# Atomic-site dependent pairing gap in monolayer FeSe/SrTiO$_3$(001)- (√13 × √13)

Cui Ding,[1,2] Zhongxu Wei,[3] Wenfeng Dong,[1] Hai Feng,[1] Mingxia Shi,[1] Lili Wang,[1,4] Jin-Feng Jia,[2,3,5] and Qi-Kun Xue[1,2,3,4]

[1]State Key Laboratory of Low-Dimensional Quantum Physics, Department of Physics, Tsinghua University, Beijing 100084, China

[2] Quantum Science Center of Guangdong-HongKong-Macao Greater Bay Area, Shenzhen 518045, China

[3]Department of Physics, Southern University of Science and Technology, Shenzhen 518055, China

[4]Frontier Science Center for Quantum Information, Beijing 100084, China

[5]Department of Physics and Astronomy, Shanghai Jiao Tong University, Shanghai 200240, China

[#]Corresponding authors: liliwang@tsinghua.edu.cn, jfjia@sjtu.edu.cn, qkxue@tsinghua.edu.cn

The interfacial FeSe/TiO$_{2-\delta}$ coupling induces high-temperature superconductivity in monolayer FeSe films. Using cryogenic atomically resolved scanning tunneling microscopy/spectroscopy, we obtained atomic-site dependent surface density of states, work function, and pairing gap in the monolayer FeSe on SrTiO$_3$(001)-(√13 × √13)-R33.7° surface. Our results disclosed the out-of-plane Se-Fe-Se triple layer gradient variation, switched DOS for Fe sites on and off TiO$_{5\square}$, and inequivalent Fe sublattices, which gives global spatial modulation of pairing gap contaminant with the (√13 × √13) pattern. Moreover, the coherent lattice coupling induces strong inversion asymmetry and in-plane anisotropy in the monolayer FeSe, which is demonstrated to correlate with the particle-hole asymmetry in coherence peaks. The strong atomic-scale correlations in lattice and electronic structure, and pairing gap in particular, put constraints on exploring the unconventional high-temperature superconductivity emerging from interface coupling, e.g., strong demand for atomic-scale interface engineering and characterization.

The monolayer FeSe interfaced with dual-TiO$_{2-\delta}$ layer is a prototypical high-temperature superconducting heterostructure in the 2D limit, particularly an exclusive charge carrier layer that can be directly probed (*1, 2*). Previous spectroscopy investigations revealed that the pairing gap varies with the lattice and reconstructions in the TiO$_{2-\delta}$ adlayer (*3-5*), exhibits micro-scale gradient tuned by oxygen vacancies and even nano-scale inhomogeneity (*6, 7*), suggesting delicate interface coupling. Compared with bulk FeSe, the quasi-two-dimensional FeSe$_4$ tetrahedra undergoes in-plane expansion with asymmetric Se heights relative to the Fe-plane due to strong interface coupling (*8, 9*). Although significant interfacial electron transfer and electron-phonon coupling have been revealed to cooperatively contribute the enhanced superconductivity (*10-13*), the correlation between the varied coupling strength and interface structure remains elusive.

Here we chose SrTiO$_3$(001) -($\sqrt{13} \times \sqrt{13}$)-R33.7° surface as the substrate, for it hosts the most open network of adlayer TiO$_{5\square}$ among all the dual-TiO$_{2-\delta}$ terminated SrTiO$_3$(001) surfaces (*14-16*). With a relatively long TiO$_{5\square}$ - TiO$_{5\square}$ separation comparable to the coherence length of bulk FeSe (*17*), it acts as a promising platform to investigate the correlation between the local pairing and FeSe$_4$/TiO$_{5\square}$ coupling. Using the cryogenic scanning tunneling microscopy/spectroscopy, we collected topographic morphology, tunneling conductance, and work function of the ($\sqrt{13} \times \sqrt{13}$) surface and the sequential epitaxial monolayer FeSe, and disclosed atomic-site varied paring gaps, manifested by varied gap magnitude and coherence peaks in anticorrelation, that can be classified into inequivalent Fe sublattices and switch on/off TiO$_{5\square}$. Our results demonstrate the intimate correlation between local pairing and the intertwined lattice and electronic modulation that emerged from the interface coupling and reveal the essential role of structure symmetry.

**1 Twofold symmetry and dipole in the TiO$_{2-\delta}$ adlayer**

As depicted in Fig. 1(a), the ($\sqrt{13} \times \sqrt{13}$) surface of the dual-TiO$_{2-\delta}$ terminated surface consists of truncated TiO$_{5\square}$ octahedron pentameters in Z-pattern stacked on the base layer of bulk-like TiO$_6$ octahedrons (*14-16*). Figure 1(b) displays the atomically resolved topographic image of the ($\sqrt{13} \times \sqrt{13}$) reconstructed surface, showing checkerboard patterns of dark and faint dark tiles, corresponding to the large vacant (LV) and small vacant (SV) zones in 1(a), respectively, separated by bright borders of Z-patterned TiO$_{5\square}$ pentameters. The inserted fast Fourier transformation (FFT) image shows two sets of Bragg spots corresponding to the two coordinate units A$_0$/B$_0$ and A$_0$'/B$_0$' marked in Fig. 1(a), with broken rotational symmetry (A$_0$ = 1.35 nm and B$_0$ = 1.44 nm) resulting from the in-plane lattice anisotropy a$_0$ < b$_0$ due to low-temperature antiferroelectric distortion in bulk SrTiO$_3$ (*18*). We notated the A$_0$/B$_0$ borders with z-pentamers marked in purple/cyan as short/long borders (SB/LB), and the corresponding Z-pentamers as SZ/LZ zones. The scattered bright dots on the borders, marked by green solid lines, correspond to equatorial oxygen vacancies O$_e$ at the inner corner sites of the Z-pentamers, which induces local elongation of ~ 0.05 nm.

The adlayer TiO$_{5\square}$ octahedrons act as electropositive species, creating positive surface dipole and lowering surface work function in turn; while O$_e$ vacancies induce further lowering. Displayed in Fig. 1(c) is the work function $\varphi$ mapping image of the same region in Fig. 1(b), and Fig. 1(d) the zoom-in image of the regular region marked by the white squares, and the statistics summarized in Table 1 (details in SM Fig. S1). The O$_e$ vacancies induce local minimum (yellow dashed lines), averagely 0.4 eV lower, but with a spatial offset of ~ a$_0$, suggesting delocalized electrons. In the regular regions without O$_e$ vacancies (Fig. 1(d)), the LV center (the apical oxygen O$_a$ of the base-layer TiO$_6$ octahedrons) shows the highest work function value, and the SZ and LZ zones show averagely comparable values of 0.10 eV lower, but the SZ zone exhibit stronger spatial variation. Similar to the case of O$_e$ vacancies, there exist unidirectional offsets of 0.5a$_0$ for local minimum. Resulting from the delocalization, the SV zone has an average value of 0.05 eV lower than SZ/LZ zones. The unidirectional feature hints at long-range repulsion accompanied by antiferroelectric rotations, which deserves further study.

**2 Coherent coupling between FeSe$_4$ and TiO$_{5\square}$**

Figure 1(e) depicts the interface stacking resolved from scanning transmission electron microscopy and total-reflection high-energy positron diffraction measurements (*8, 9*). The bottom-layer Se atoms reside at the hollow sites of the topmost TiO$_{2-\delta}$ adlayer, which are closer to the Fe layer than the upper-layer Se atoms, i.e., $h_1$(Se-bottom) < $h_2$(Se-upper). The broken inversion symmetry, together with broken in-plane rotational symmetry, gives inequivalent Fe sublattices, i.e., α-Fe and β-Fe with flipped [100]/[010] direction bonding with bottom/upper layer Se dimers. Thus, the four zones contain various Fe atoms, that is, two/six α-Fe and two/six β-Fe atoms off TiO$_{5\square}$ in the SV/LV zone and three/two α-Fe and two/three β-Fe atoms on TiO$_{5\square}$ in the SZ/LZ zone.

The monolayer FeSe exhibits concomitant atomic contrast with the (√13 × √13) reconstruction, exactly, the Z-TiO$_{5\square}$-pentamer. The bias-dependent atomically resolved images of monolayer FeSe displayed in Figs. 1(f)-(h) consistently present expanded Se-terminated (001) lattices with particularly strong atomic contrast under small sample bias (Figs. 1(f,g), $V_s$ = ± 50 meV), indicative of tuned Fe-3$d$ orbitals upon TiO$_{5\square}$. The paired apparent bright and long borders, with purple marks, show strong atomic contrast, while the other paired faint borders marked in cyan exhibit negligible atomic contrast. Reminding the contrasting spatial variations in the work function image, we infer those with strong atomic contrast to the SZ zones, that is, corresponding to α-Fe-chains along the smaller lattice $a_0$ and, thus, under weaker expansion. A detailed investigation on the nearest neighbor Se-Se separation (SM Fig. S2) reveals smaller (larger) separation for Se-dimers on single TiO$_{5\square}$ than those either bridged two TiO$_{5\square}$ or with single Se atom on TiO$_{5\square}$ at 50 meV (-50 and -250 meV), and opposite lattice anisotropy (γ=a/b) on and off TiO$_{5\square}$ pentamer (e.g., γ > 1 on Z-zones vs. γ <1 on vacant zones under small bias of ± 50 meV, and reverse at a large bias of 250 meV). The above bias- and TiO$_{5\square}$-dependent features indicate electronic modulations derived by the coherent FeSe$_4$/TiO$_{5\square}$ lattice coupling, which might share similar orbital-dependent quasiparticle coherence as disclosed in bulk FeSe (*19*).

## 3 Electronic and pairing modulation in the monolayer FeSe

### 3.1 Se-Fe-Se triple layer dependence and inequivalent Fe sublattices

The coherent FeSe/TiO$_{2-\delta}$ lattice coupling induces the modulation of electronic structure and superconductivity in the monolayer FeSe with spatial variation upon single Z-pentamer and single TiO$_{5\square}$, as further resolved from the spectroscopy measurement of work function $\varphi$ displayed in Fig. 1(i) and tunneling conductance $g$ (r, E) = d$I$/d$V$ (r, V) in Figs. 2(b,c). The work function mapping image in Fig. 1(i) and the typical FFT image in Fig. 1(j) show the same periodicity as (√13 × √13) units but with opposite contrast to the topography images in Figs.

1(f-h), i.e., higher work function values at the LV center and in the LZ zone. The statistics by category of two Se layers and two Fe sublattices give a gradual increase towards the surface and relatively small value in the α-Fe, i.e., $\varphi$(Se-bottom) < $\varphi$(α-Fe) < $\varphi$(β-Fe) < $\varphi$(Se-upper) (Table 1 and SM Fig. S2), indicating interface positive dipole and inequivalent Fe sublattices. On the other hand, the averaged work function value of Fe-sites in the four divided zones follows the sequence $\varphi$(SV) < $\varphi$(SZ) < $\varphi$(LZ) < $\varphi$(LV), agreeing well with the trend on SrTiO$_3$(001)-($\sqrt{13} \times \sqrt{13}$) surface and the comparison $\varphi$(α-Fe) < $\varphi$(β-Fe) (for flipped α-Fe and β-Fe combination in SZ and LZ zones).

The typical large-bias tunneling spectra in Fig. 2(b) exhibit steplike features, including low tunneling conductance around $E_F$ ranging from -50 meV to 60 meV from the M electron band (inferred as δ, Fe-$d_{xy}$/$d_{yz/xz}$ orbital dominant), upturn at a negative bias of about -80 meV from the onsets of Γ hole band (σ, Fe-$d_{yz}$/$d_{xz}$ orbital dominant), upturn at a positive bias of about 90 meV from the Γ electron band (η, Fe-$d_{xy}$ orbital dominant), and hump at -240 meV (ξ, Fe-$d_{3z^2-r^2}$ orbital dominant) (20-22). The typical small-bias tunneling spectra in Fig. 2(c) exhibit two-gap features, i.e., inner-gap with coherence peaks at $\Delta_i$ = ± 10 meV (black dashed lines), and outer-gap with coherence peaks varied around $\Delta_o$ ~ ± 17 meV (blue dashed lines) with contrasting broadening up to ± 20 meV (green dashed lines).

Within each zone, the Fe sites have narrower and flatter δ than the Se sites, correspondingly, more prominent two-gaps with smaller outer-gap magnitudes but sharper coherence peaks and wider zero-conductance plateau around $E_F$, characteristics of more spectra weight loss due to pairing, i.e., higher superfluid density. The β-Fe exhibits stronger coherence peaks, accompanied by a little more prominent σ-band and η-band edges, whereas the α-Fe sites show widening outer gaps (green dashed lines). Moreover, the Fe sites on the middle TiO$_{5\square}$ (Fe-m) have sharper coherence peaks than those on the terminal- and corner-ones (inferred as Fe-t and Fe-c, respectively). Compared with upper-layer Se atoms, the bottom-layer Se atoms exhibit slightly lower DOS between ξ and σ bands and higher ± $V_{in}$ coherence peaks, bearing more resemblance to Fe sites. Notably, the pairing gap contrasts are more directly discernable from the normalized spectra (SM Fig. S3).

**3.2 Switch between on and off TiO$_{5\square}$**

The tunneling spectra in Figs. 2(b,c) disclose remarkable zone-dependent features. Statistically, the SZ zone (containing 2-α-Fe-t, 1-α-Fe-m and 2-β-Fe-c) exhibits the highest ξ-band energy (closer to $E_F$), the highest DOS around the Fermi level, and the largest outer-gap magnitude but the weakest coherence peaks and the less spectra weight loss of pairing. Whereas, the LZ zone (containing 2α-Fe-c, 1β-Fe-m and 2β-Fe-t) resides in the other end, and LV and SV zones (containing Fe atoms off $TiO_{5\square}$) in the middle sequence (Figs. 2(b,c)). Another striking feature is switched particle-hole asymmetric outer-gap coherence peaks ± $V_o$. The SV zone exhibits stronger negative coherence peaks - $\Delta_o$, whereas the SZ zone has stronger positive coherence peaks + $\Delta_o$, while the LZ zone exhibits the weakest particle-hole asymmetry (Fig. 2(c)).

To get further insight into the (√13 × √13)-tuned Fe-3$d$-orbitals and the corresponding pairing gaps, we collected bias-dependent tunneling conductance mapping images at the characteristic energies of Fermi level (0 V, Fig. 3(b)), coherence peaks of inner gap $\Delta_i$ (± 10 meV, Figs. 3(c) and 3(d)), around the outer gaps $\Delta_o$ (± 20 meV, Figs. 3(e) and 3(f)), δ-band (edges at ± 50 meV, Figs. 3(g) and 3(h)), η-band (80 meV, Fig. 3(i)), and σ-band (-80 meV, Fig. 3(j)). The mapping images, together with the corresponding FFT images displayed in Figs. 3(k-o) (more data in SM Fig.S4), consistently show remarkable contrast with robust unidirectionality. The LZ and SV zones exhibit the strongest ± $V_i$ coherence peaks and the highest DOS outside - $V_o$, associated with similar contrast in the σ-band and the negative edge of the δ-band. The SZ zone exhibits the lowest DOS outside -$V_o$ but the highest DOS outside + $V_o$, associated with the lowest DOS at σ-band and the negative edge of δ-band. Whereas, the LV zone is right reverse. The above unidirectional spatial modulations demonstrate the universal switched DOS contrasts in Fe-sublattices between on and off $TiO_{5\square}$. As directly identified from the DOS difference $g(\alpha\text{-Fe})-g(\beta\text{-Fe})$ presented in Fig.4(d), the α-Fe has higher DOS around $E_F$ (within ± 30 meV) but converts to lower at higher energy in the Z-pentamer zones, whereas, right reverse in the vacant zones.

**3.3 Correlation with in-plane lattice and symmetry**

Besides bias-switched electronic anisotropy from topographic images, the in-plane asymmetry is further identified from tunneling conductance mapping images. The intensity contrast between the $q_A$ and $q_B$ Bragg spots is $I_{qA} > I_{qB}$ at 0 -10, and -20 meV but $I_{qA} < I_{qB}$ at larger energies.

The unidirectional electronic structure are directly identified from the respective inverse-FFT images in Fig. S4. To figure out the correlation with electronic symmetry, we compared two adjacent (√13 × √13) units with contrasting in-plane anisotropy, as depicted in Fig. 4(a). The Se-Se separation comparison (Fig. S5) indicates averagely less expanded Se(001) lattices with stronger anisotropy above Z-pentamers but weaker anisotropy off Z-pentamers in unit II. We notated the Fe sites in unit II as α'-Fe and β'-Fe to differentiate the local lattice anisotropy.

With the distinct local anisotropy, the two units indeed exhibit varied contrast between Fe sublattices and on/off $TiO_{5\square}$, despite consistent dipole-induced Se-Fe-Se triple layer gradient (Fig. S5). Figures 4(b) and 4(c) summarize the typical large bias and small bias d$I$/d$V$ tunneling spectra taken on Fe sites in the two units, respectively, and Fig. 4(d) the conductance difference between Fe sublattices. Compared with unit I, unit II exhibits lower ξ-bands, lower DOSs between ξ- and σ-band, higher DOSs above η-band and lower work function (SM Fig. S5), associated with softer paring gap characterized by weaker coherence peaks and lowered spectra weight loss around $E_F$, and narrower zero DOS plateau at $E_F$. The DOS contrast between α'-Fe and β'-Fe switch between the Z-pentamer and vacant zones with similar bias limits of ± 30 meV. However, the contrast around $E_F$ is stronger (larger deviation from zero within ± 30 meV) in the LZ zone and weaker in the vacant zones (except for the remarkable difference around coherence peaks). Correspondingly, the Fe sites in the vacant zones have relatively harder pairing gaps, in sharp contrast to the outstanding β-Fe-m in the LZ zone of unit I. The above contrasting features consistently indicate the essential role of local in-plane symmetry, that is, the more expanded and symmetric in-plane lattices (associated with the more symmetric electronic structure), the larger and harder the pairing gaps.

**Conclusion and summary**

Our spatially atomically resolved cryogenic STM/STS characterization on the FeSe monolayer films on $SrTiO_3$(001)-(√13 × √13) surface consistently demonstrate atomic-scale modulations in topography (Figs. 1(f-h)), DOS around $E_F$ (Figs. 2(b) and 4(b)), work function (Fig. 1(i)), and pairing gap (Figs. 2(c) and 4(c)) in particular, resulting from strong interfacial FeSe/$TiO_{2-\delta}$ coupling. The universal atomic-scale modulations originate from strong lattice-electronic interaction with intertwined in-plane and out-of-plane lattice asymmetry in the flexible FeSe$_4$ tetrahedra under tensile strain. The in-plane tensile strain, associated with reduced Se height, thus, enhanced two-dimensionality, gives the unique Fermi structure of the monolayer FeSe, featured by flattened δ-band with upward-shifted σ-band and ξ-band, instead of the downward shifted bands under electron-doping (*23*).

The FeSe/TiO$_{2-δ}$ coupling introduces charge transfer following the band-bending scenario in semiconducting interfaces(*24*) with atomic-layer gradation under the interface dipole effect. This is evidenced by the gradient-increased work function values from the bottom Se(001) to Fe(001) and then to the upper Se(001) layer. On the other hand, in-plane modulations follow the TiO$_{5□}$ Z-pentamer motifs on the SrTiO$_3$(001) surface, as evidenced by both work function (Table 1) and DOS (Fig. 4(d)) switch between on and off TiO$_{5□}$ pentamers. Notably, the work function is deduced from the tunneling current dominantly contributed by the Γ-centered bands, for the much higher tunneling probability than the M-centered δ electron bands (which dominate the DOS at $E_F$) (*20*). Thus, the work function values of monolayer FeSe are much higher than the values of about 4.5 eV measured by X-ray photoemission spectroscopy (*25*), and the spatial varied work function might lack a direct correlation with the pairing gap opened at the M-centered electron pockets (*26*). The relatively larger work function coincides with the sharper upturns from the Γ-centered bands and relatively low tunneling conductance around $E_F$ (β-Fe vs. α-Fe in Fig. 2(b) and unit I vs. unit II in Fig. S5, supporting enhanced 2D electronic structure.

The most important finding is the paring gaps tuned by the TiO$_{5□}$-Z-pentamer exhibit anti-correlated gap magnitude and coherence peaks and switched particle-hole asymmetry. The out-of-plane electronic dipole and in-plane lattice asymmetry cooperatively affect the particle-hole symmetry, as exemplified by the remarkable contrasts in on/off TiO$_{5□}$-Z-pentamer, Se-Fe-Se triple layers, and Fe sublattices ((Figs. 2(c) and 4(c))). Despite delocalized charge transfer, the coherent FeSe$_4$/TiO$_{5□}$ lattice delicately tuned the spectra weight transfer, as evidenced by the universal contrasts in the nearest neighbored α-Fe and β-Fe and the fine variation with the sequence of single TiO$_{5□}$ (Fe-m vs. Fe-t vs. Fe-c), wherein the symmetry plays an essential role. Though the tunneling current detected by the atomic sharp tip removes the momentum resolution due to the Heisenberg uncertainty principle, the bias-dependent reversed features in both DOS contrast and the in-plane anisotropy suggest orbital-selective tunneling probability. The robust contrast between α-Fe and β-Fe puts the inequivalent Fe sublattices as competitive atomic-scale lattice carriers of electronic nematicity.

In summary, we disclosed atomic-site dependent pairing gap in monolayer FeSe tuned by intertwined lattice and electronic interaction under broken inversion and rotational symmetry. The atomic-scale FeSe/TiO$_{2-\delta}$ coupling and its correlation with Fermi structure and pairing disclosed the essential role of 2D lattice and symmetry. The anti-correlated gap magnitude and coherence peaks and switched particle-hole asymmetry observed here are ubiquitous features in cuprates(*27-29*). Our results put constraints on understanding the complex features in high-temperature superconductivity emerging from heterostructures, for example, pseudogap and nematicity. On the other hand, the robust broken inversion symmetry in the quasi-2D FeSe monolayer provides a promising platform to investigate the spin-orbital coupling effect.

**Materials and Methods**

Our experiments are carried out in a Createc ultrahigh vacuum (UHV) low-temperature STM system equipped with a molecular beam epitaxy (MBE) chamber for film preparation. The base pressure is better than $1.0 \times 10^{-10}$ Torr. The Nb-doped SrTiO$_3$(001) (0.05 wt. %) substrates were annealed at 880 °C for 2-3 hours to obtain atomically flat TiO$_{2-\delta}$-terminated surface with ($\sqrt{13} \times \sqrt{13}$) reconstruction. FeSe films were then grown by co-evaporating high-purity Fe (99.995%) and Se (99.999%) from standard Knudsen cells at a substrate temperature of 380 °C. The K-cell temperatures of Fe and Se were 1010 and 90 °C, respectively, corresponding to a Fe/Se flux ratio of ~ 1:10 and a deposition rate of ~ 0.02 unit cell per minute. At last, the samples were annealed at ~ 450 °C for 1 hour to remove excess Se atoms.

All STM measurements were performed in a constant current mode at 4.8 K, with a polycrystalline PtIr tip and the bias voltage applied to the sample ($V_s$). The tunneling current set point $I_t$ is 100 pA for the SrTiO$_3$(001) surface and 500 pA for the monolayer FeSe. The differential conductance d$I$/d$V$ spectra, characterizing the local density of states around the Fermi level, are measured by disabling the feedback circuit, sweeping the sample voltage $V_s$, and then extracting the differential tunneling current d*I*/d*V* using a standard lock-in technique with a small bias modulation (~ 1 % of the sweeping range) at 937 Hz. To calculate the local tunneling barrier height $\varphi = 0.95 \left(\frac{d\ln I}{dz}\right)^2$ (*I* in the units of A and *z* in the units of Å), the values of ln*I* are measured while disabling the feedback circuit and decreasing the tip-sample distance *z*, and then dln*I*/d*z* is extracted from the linear fitting of the ln*I*-*z* relation.

**Acknowledgments**

This work is supported by the National Natural Science Foundation of China (Grants No. 52388201 and 12074210), the National Basic Research Program of China (Grant No. 2022YFA1403102), and the Basic and Applied Basic Research Major Programme of Guangdong Province, China (Grant No. 2021B0301030003) and Jihua Laboratory (Project No. X210141TL210).

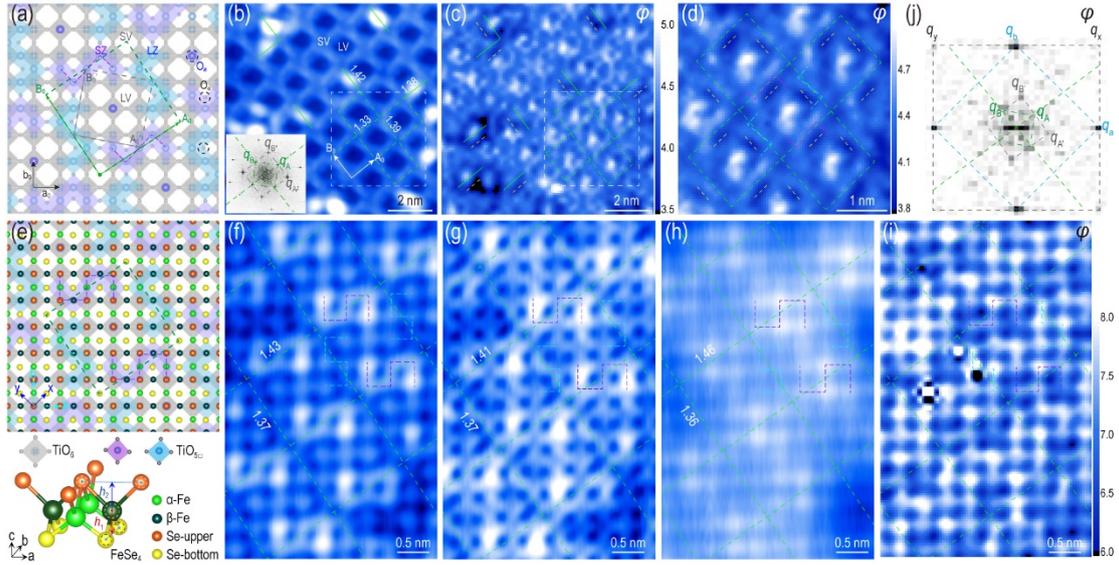

Fig.1 Topography and work function characterization of SrTiO₃(001)-(√13 × √13) surface and monolayer FeSe films thereon. (a) Schematic illustration of the dual-TiO$_{2-\delta}$ termination of SrTiO₃(001)-(√13 × √13) surface. (b) Topographic image ($V_s$ = 1 V), with typical Fast Fourier transformation (FFT) pattern inserted, and (c) work function mapping image ($V_s$ = 1V) of the same region on SrTiO₃(001)-(√13 × √13) surface, and (d) zoom-in work function image of the area marked by the white dashed square in (b,c). (e) Schematic illustration of FeSe₄ tetrahedra and interface stacking on the dual-TiO$_{2-\delta}$. (f-h) Bias-dependent atomically resolved topographic images (f, $V_s$ = 50 mV; g, $V_s$ = -50 mV; h, $V_s$ = -250 mV) of the monolayer FeSe, and (i) work function mapping image ($V_s$ = 50 mV) of the same region. (j) Typical FFT pattern of the work function image of monolayer FeSe.

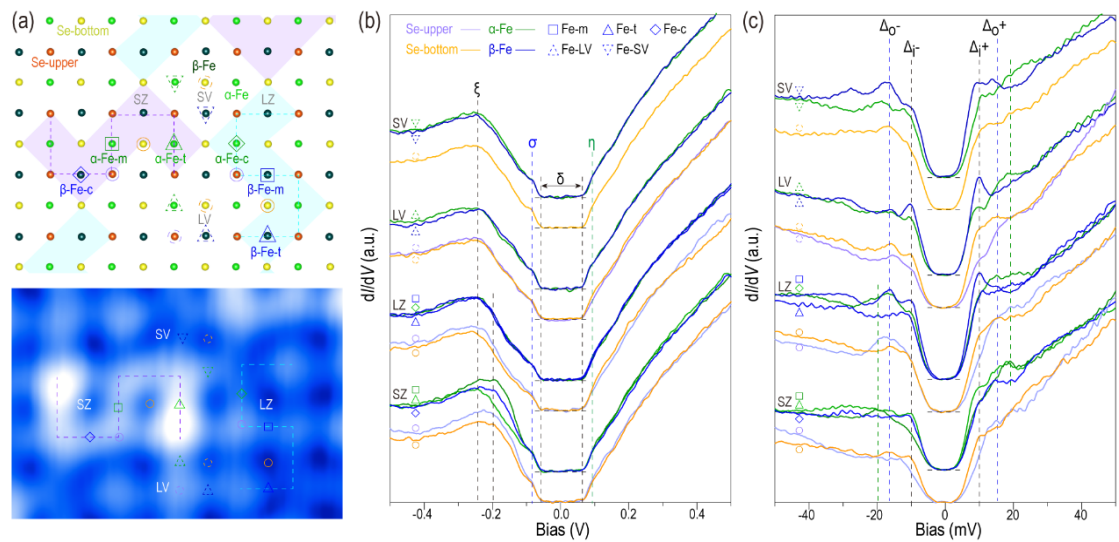

Fig. 2 Atomic-site dependent tunneling spectra of monolayer FeSe tuned by the (√13 × √13) pattern. (a) Schematic illustration of atomic sites and atomically resolved topographic image ($V_s$ = 50 mV) of monolayer FeSe on SrTiO$_3$(001)-($\sqrt{13} \times \sqrt{13}$). (b) Large bias ($V_s$ = 500 mV) and (c) small bias ($V_s$ = 50 mV) d$I$/d$V$ tunneling spectra of monolayer FeSe taken at various points marked in (a). The spectra are offset for clarity, with horizontal dashed lines indicating the zero-conductance. The dashed lines in (b) are eye-guides for the labeled bands, and the black, blue, and green ones in (c) mark the coherence peaks for inner gap $\Delta_i$ at ± 10 meV, outer gap $\Delta_o$ at ± 17 meV and broadening up to ± 20 meV, respectively.

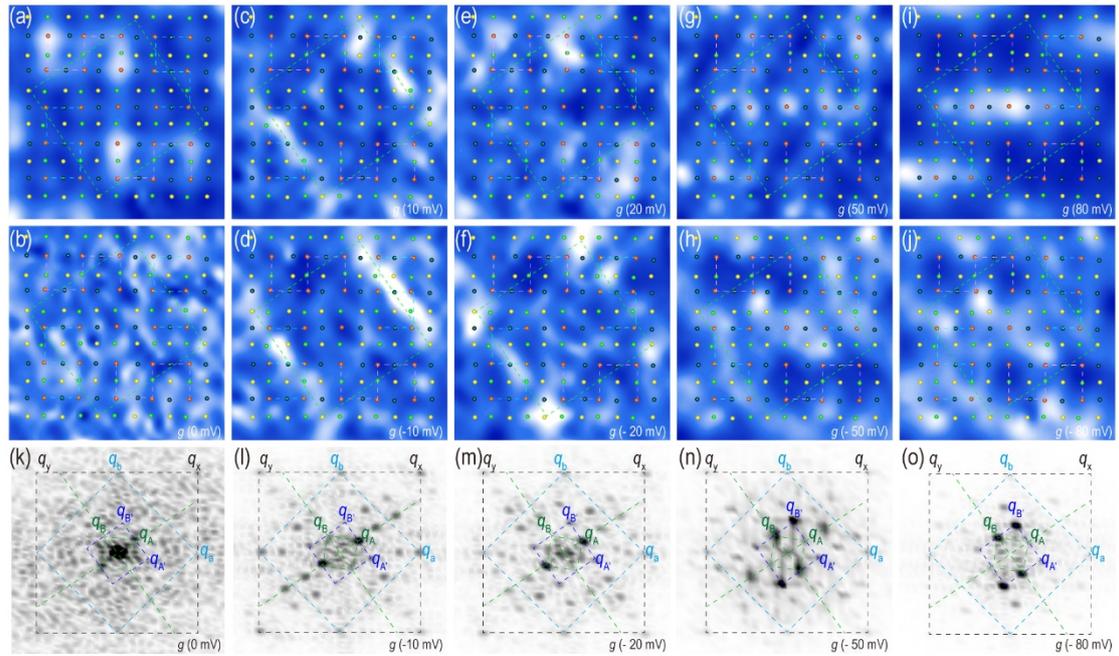

Fig. 3 Bias-dependent d$I$/d$V$ mapping images of monolayer FeSe on SrTiO$_3$(001)-($\sqrt{13} \times \sqrt{13}$). (a) Topographic image ($V_s$ = 50 mV) and (b-j) bias-dependent d$I$/d$V$ mapping images of the same region. The green squares, purple, and cyan dashed lines mark the ($\sqrt{13} \times \sqrt{13}$) unit cells and the z-pentamers along the two perpendicular frames, respectively. (k-o) Typical bias-dependent FFT patterns of d$I$/d$V$ mapping images.

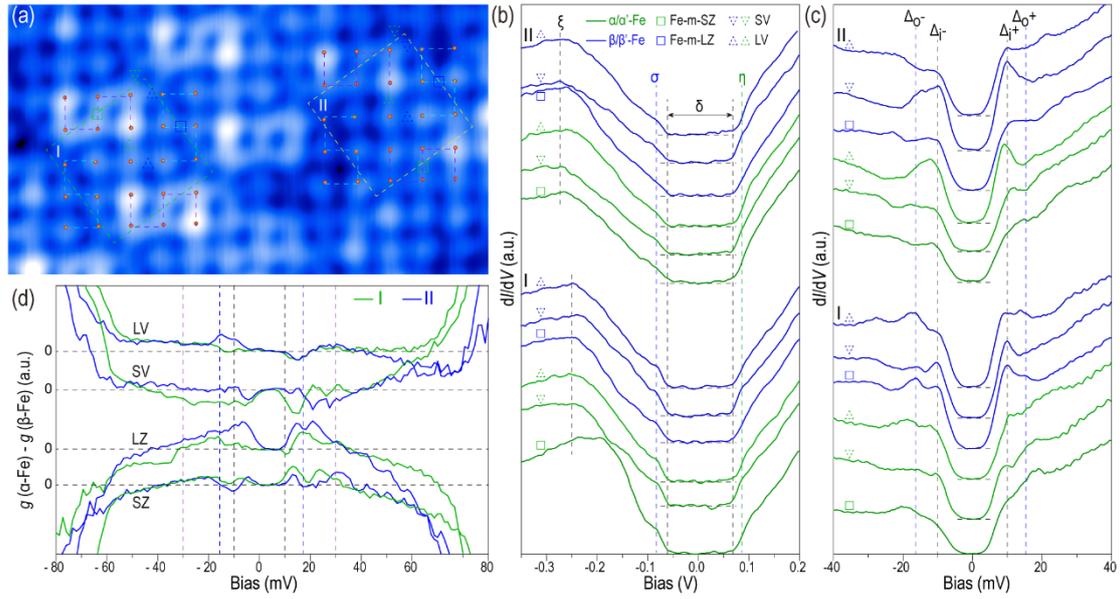

Fig. 4 Comparison of atomic-site dependent tunneling spectra of monolayer FeSe with distinct local symmetry. (a) Topographic image ($V_s$ = 50 mV) of monolayer FeSe with contrasting surface structure in the two (√13 × √13) unit cells marked in green and yellow. The red dots mark the upper layer Se atoms. (b) Large bias ($V_s$ = 500 mV) and (c) small bias ($V_s$ = 50 mV) d$I$/d$V$ tunneling spectra taken at various points marked in (a). (d) Tunneling conductance difference between α-Fe and β-Fe in the respective four zones of the two units. The black and blue dashed lines are eyes-guides for inner gap $\Delta_i$ ~ ± 10 meV and outer gap $\Delta_o$ ~ ± 17 meV, and the violet ones at ± 30 meV for the reverse contrast across the zero points.

| | $\varphi$ (eV) | | SV | LZ | SZ | LV | With $V_O$ | Without $V_O$ |
|---|---|---|---|---|---|---|---|---|
| SrTiO$_3$(001) | $\varphi_{averaged}$ | | 4.29 ± 0.11 | 4.33 ± 0.09 | 4.34 ± 0.17 | 4.43 ± 0.22 | 3.95 ± 0.29 | 4.33 ± 0.19 |
| | $\varphi_{max\text{-}Gaussian}$ | | 4.28 ± 0.01 | 4.32 ± 0.01 | 4.35 ± 0.02 | 4.41 ± 0.03 | 3.96 ± 0.02 | 4.31 ± 0.01 |
| FeSe/SrTiO$_3$(001) | $\varphi_{Se\text{-}bottom}$ | $\varphi_{averaged}$ | 6.40 ± 0.20 | 6.62 ± 0.18 | 6.36 ± 0.23 | 6.53 ± 0.22 | α-Fe | β-Fe |
| | | $\varphi_{max\text{-}Gaussian}$ | / | 6.45 ± 0.05 | 6.25 ± 0.07 | 6.52 ± 0.03 | $\varphi_{averaged}$ | |
| | $\varphi_{Fe}$ | $\varphi_{averaged}$ | 7.14 ± 0.30 | 7.41 ± 0.22 | 7.19 ± 0.26 | 7.34 ± 0.33 | 7.27 ± 0.33 | 7.31 ± 0.18 |
| | | $\varphi_{max\text{-}Gaussian}$ | / | 7.40 ± 0.01 | 7.27 ± 0.02 | 7.36 ± 0.03 | $\varphi_{max\text{-}Gaussian}$ | |
| | $\varphi_{Se\text{-}upper}$ | $\varphi_{averaged}$ | / | 7.85 ± 0.19 | 7.59 ± 0.32 | 8.01 ± 0.14 | 7.31 ± 0.05 | 7.32 ± 0.02 |
| | | $\varphi_{max\text{-}Gaussian}$ | / | 7.82 ± 0.03 | 7.43 ± 0.03 | 7.95 ± 0.09 | | |

Table 1 Statistical work function values of SrTiO$_3$(001) and FeSe/SrTiO$_3$(001) deduced from the statistics results in Figs. S1(d,e) and Fig. S2(j-m).

# Supplementary Materials

## Atomic-site dependent pairing gap in monolayer FeSe/SrTiO$_3$(001)- (√13 × √13)

Cui Ding, Zhongxu Wei, Wenfeng Dong, Hai Feng, Mingxia Shi, Lili Wang, Jin-Feng Jia, and Qi-Kun Xue

**Gaussian Fitting on Statistics of Work Function**

We applied Gaussian function $y = y_0 + Ae^{-0.5\left(\frac{x-x_c}{w}\right)^2}$ to fit the statistical distributions of work function displayed in Figs. S1(d,e) and Fig. S3. The $\varphi_{max-Gaussian}$ in table I corresponds to the value $x_c$ and the corresponding error is the standard deviation of the Gaussian distribution.

**Mapping image interpolation**

To better show the spatial distribution of electronic states, the bias-dependent d$I$/d$V$ mapping images presented in Fig. 3 is obtained by interpolating the zoom-in raw data (Fig. S4) with linear interpolation method.

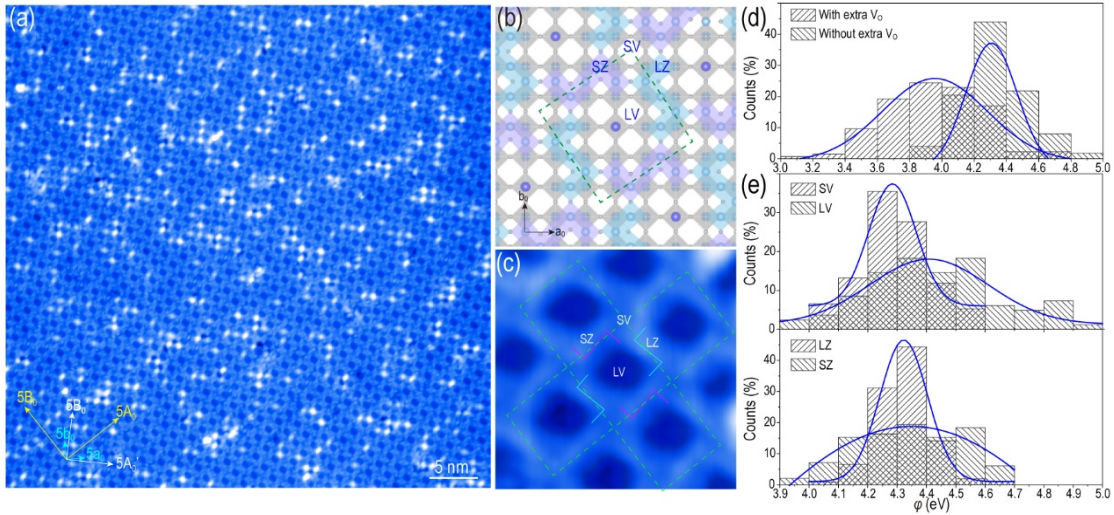

Fig. S1 (a) Topographic image ($V_s$ = 1 V, $I_s$ = 50 pA) of SrTiO$_3$(001)-(√13 × √13) surface. The corresponding FFT pattern is inserted in Fig. 1(b). (b) Schematic illustrations of SrTiO$_3$(001)-(√13 × √13) surface. (c) Topographic image ($V_s$ = 1 V, $I_s$ = 100 pA) of the region in Fig. 1(d). The blue dashed lines mark the (√13 × √13) unit cells. (d, e) Statistical comparison of work function values between regions with/without extra oxygen vacancies (d) and between SV/LV and LZ/SZ zones without extra oxygen vacancies (e).

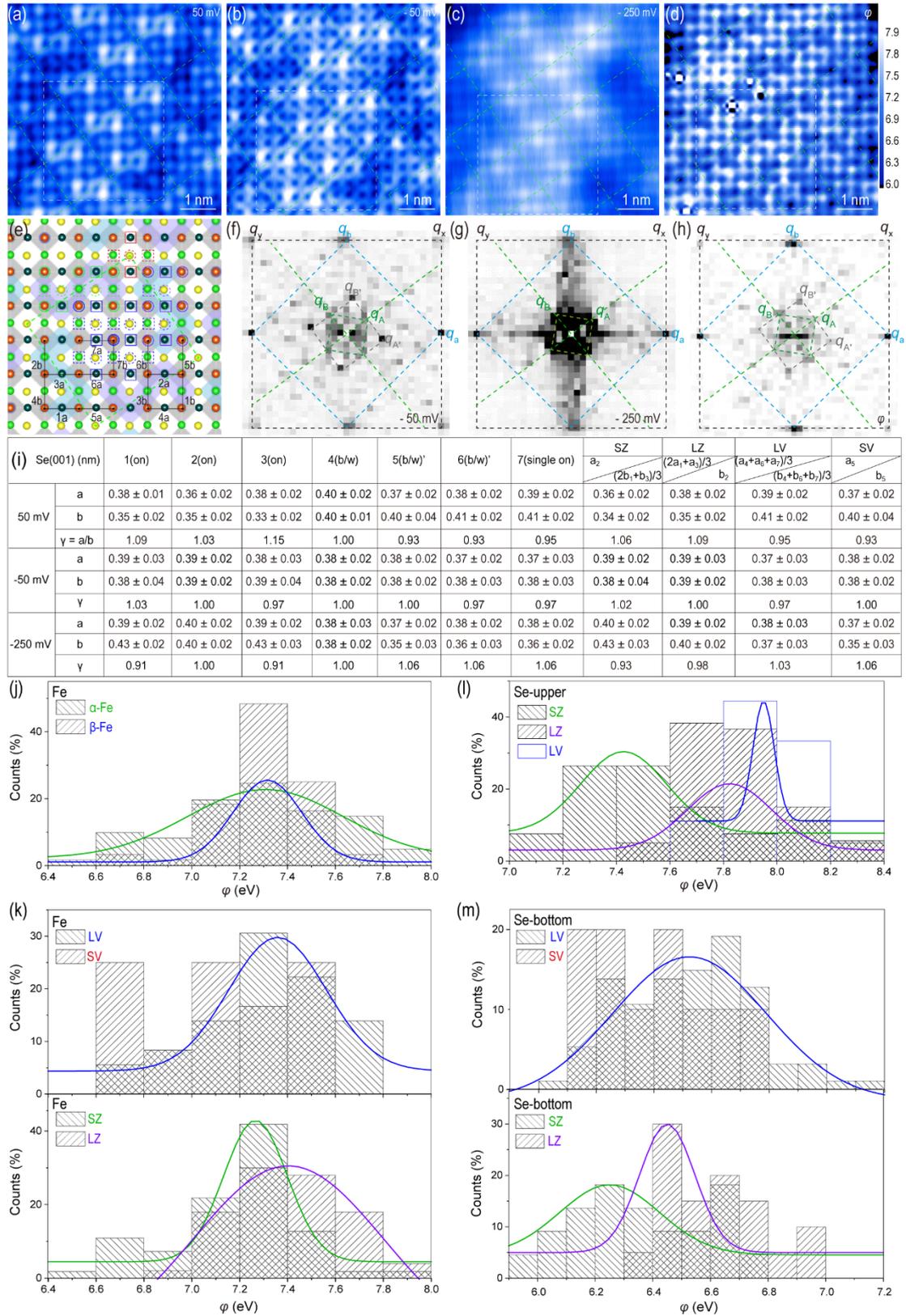

Fig. S2 (a-c) Bias-dependent topographic images and (d) work function mapping image of the monolayer FeSe on SrTiO$_3$(001)-($\sqrt{13} \times \sqrt{13}$). Set point: (a,d), $V_s$ = 50 mV; (b), $V_s$ = -50 mV; (c), $V_s$ = -250 mV. (e) Schematic illustration of epitaxial monolayer FeSe on SrTiO$_3$(001)-($\sqrt{13} \times \sqrt{13}$). (f-h) Fast Fourier transform (FFT) patterns of (b), (c), and (d), respectively. (i) Statistics on the bias-dependent nearest neighbor Se-Se separations, as labeled in (e), of the region marked by the white dashed squares in (a-c). (j-m) Statistics on work function values in the white square in (d) by groups of different layers ((j,k) Fe(001), (l) upper layer Se (001), and (m) bottom layer Se(001)), with respective Gaussian fitting curves overlaid. The LV/SV/SZ/LZ zones are marked in blue/red/green/purple as illustrated in (e).

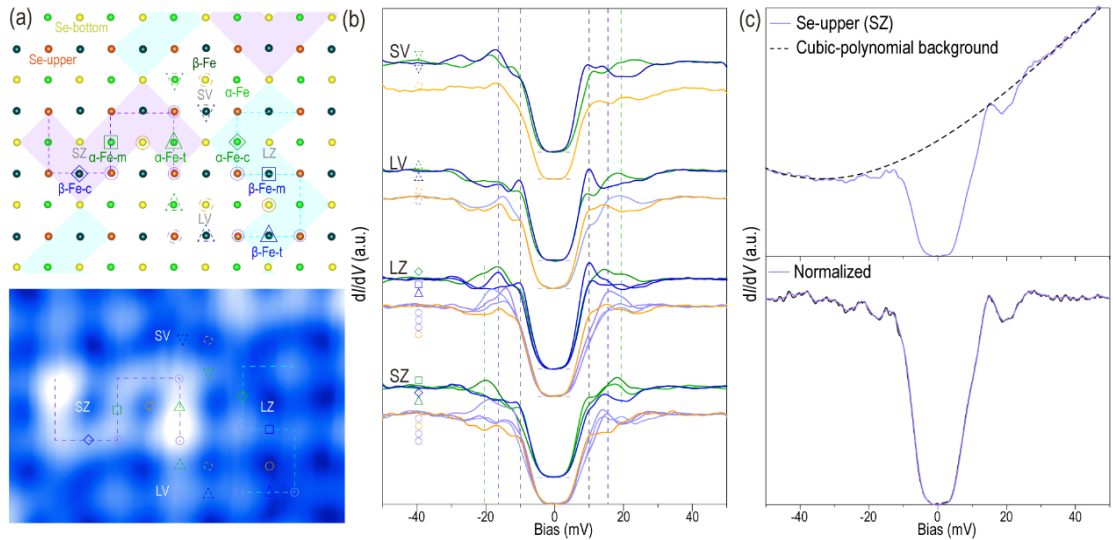

Fig. S3 Normalized and smoothed tunneling spectra of monolayer FeSe. (a) Schematic illustration of atomic sites and atomically resolved topographic image of monolayer FeSe on $SrTiO_3(001)$-($\sqrt{13} \times \sqrt{13}$). (b) The normalized-and-smoothed tunneling spectra taken at the points labeled in (a). (c) Exemplification of normalization and smoothing. Upper panel: d$I$/d$V$ spectrum taken at an upper-layer Se site in SZ zone and its cubic-polynomial background obtained by fitting to the spectrum for bias |$V$| > 30 mV (black dashed curve). Lower panel: the normalized d$I$/d$V$ spectrum by its cubic-polynomial background (black solid curve) and the smoothed spectrum by cubic-polynomial fitting (purple solid curve).

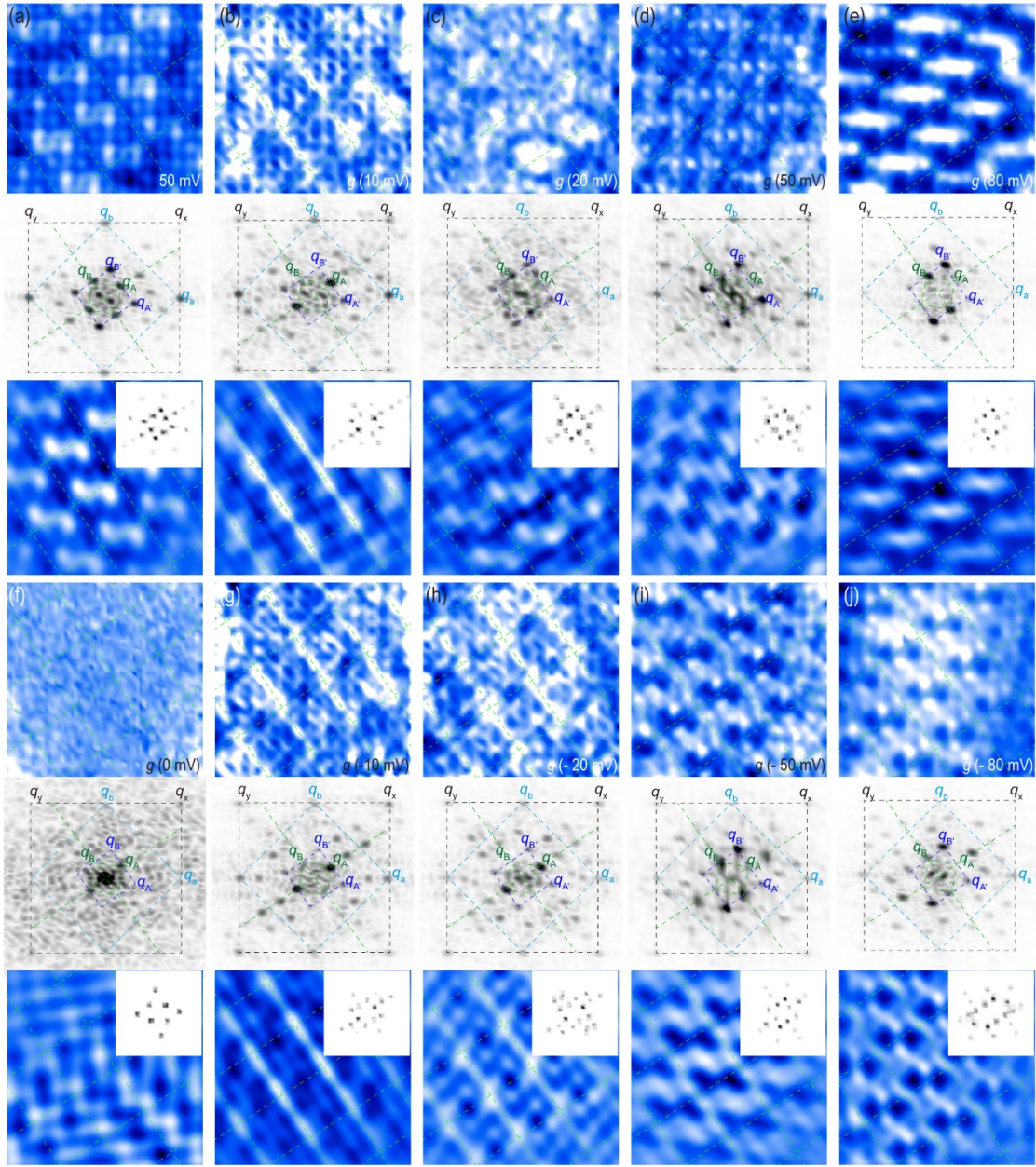

Fig. S4 (a) Topographic image ($V_s$ = 50 mV) and (b-j) the corresponding d$I$/d$V$ mapping images of the monolayer FeSe on SrTiO$_3$(001)-($\sqrt{13} \times \sqrt{13}$), with the respective FFT patterns in the middle and inverse-FFT images displayed in the bottom panels.

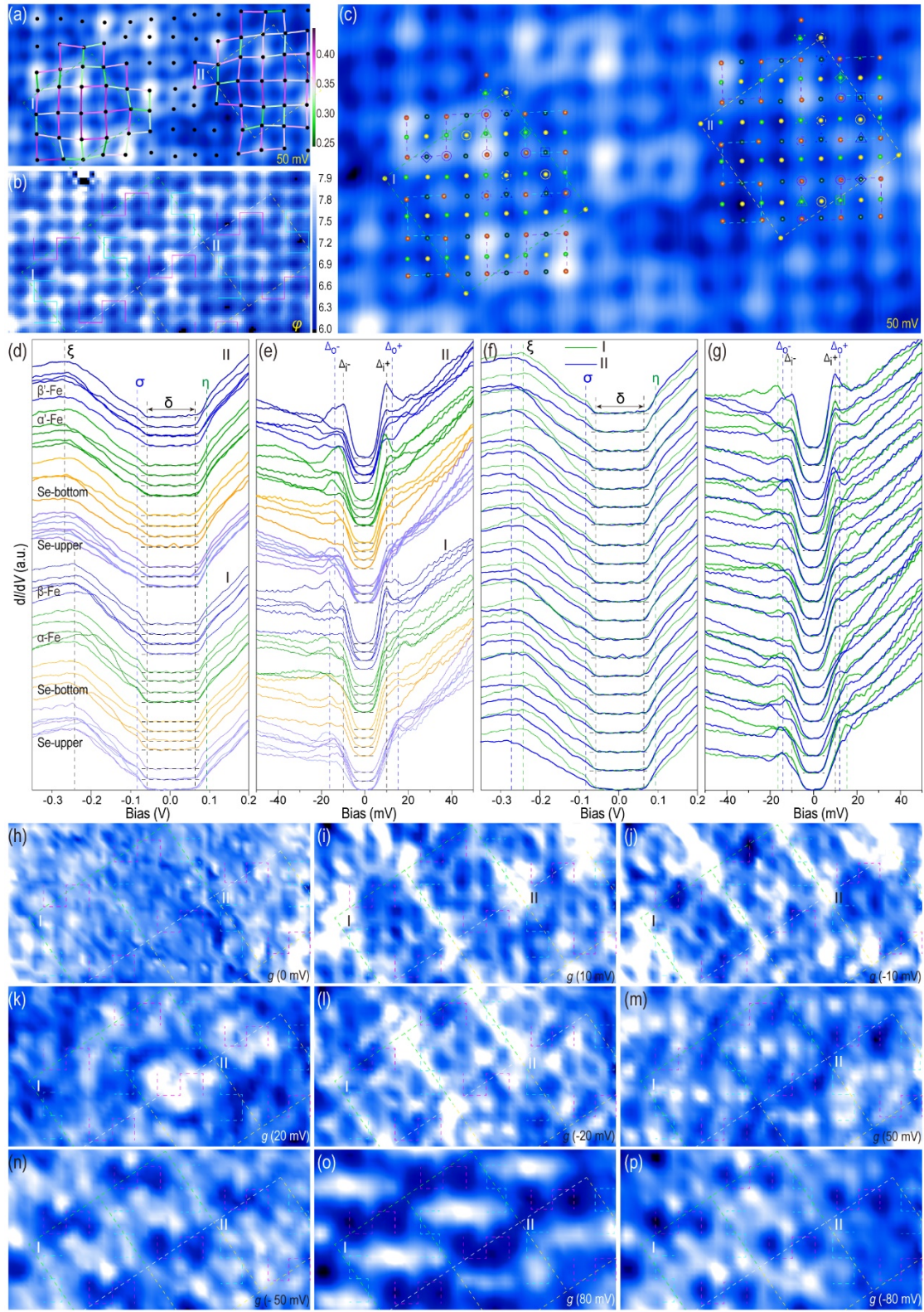

Fig. S5 (a) Topographic image, (b) work function mapping image, and (h-p) d$I$/d$V$ mapping images of monolayer FeSe in the same region. The green and yellow dashed rectangles mark the (√13 × √13) unit cells with distinct surface structures. (d,f) Large bias and (e,g) small bias d$I$/d$V$ tunneling spectra taken at the various points labeled in (c). The spectra are offset for clarity with horizontal dashed lines indicating zero-conductance positions. In (a), local maxima (black dots) are used as approximate positions of the Se atoms; the distances between adjacent atoms are manifested by the colored segments. Set point: (a-d, e, g, h-p) $V_s$ = 50 mV; (d,f) $V_s$ = 500 mV;.